## 4.7 Eigenfactor

Grischa Fraumann, Jennifer D'Souza, and Kim Holmberg

**Abstract:** The Eigenfactor™ is a journal metric, which was developed by Bergstrom and his colleagues at the University of Washington. They invented the Eigenfactor as a response to the criticism against the use of simple citation counts. The Eigenfactor makes use of the network structure of citations, i.e. citations between journals, and establishes the importance, influence or impact of a journal based on its location in a network of journals. The importance is defined based on the number of citations between journals. As such, the Eigenfactor algorithm is based on Eigenvector centrality. While journal based metrics have been criticized, the Eigenfactor has also been suggested as an alternative in the widely used San Francisco Declaration on Research Assessment (DORA).

### Introduction

This chapter provides an overview on the Eigenfactor™, a journal metric, which was developed by Bergstrom (2007) and his colleagues at the University of Washington. They invented the Eigenfactor as a response to the criticism against the use of simple citation counts (Bergstrom, 2007). They also claimed a need for alternative metrics (West, Bergstrom and Bergstrom, 2010), which in this case should not be confused with altmetrics, which are metrics to track mentions of scholarly articles online (Priem et al., 2010).

The Eigenfactor makes use of the network structure of citations, i.e. citations between journals (Bergstrom, 2007). The citations are retrieved from Journal Citation Reports (JCR), which is a part of Clarivate Analytics' Web of Science (West, Bergstrom, and Bergstrom, 2010). The Eigenfactor is defined as a flow-based journal ranking, because it simulates the workflow of a researcher searching through journals using citation links (Bohlin et al., 2016). By doing so, it is "interpreted as a proxy for how often a researcher who randomly navigates the citation landscape accesses content from the journal" (Bohlin et al., 2016). These navigational traces, i.e. citations between journals, can be used to calculate a journal's influence (Chang, McAl-

**Grischa Fraumann**, Research Assistant at the TIB Leibniz Information Centre for Science and Technology in the Open Science Lab, Research Affiliate at the "CiMetrias: Research Group on Science and Technology Metrics" at the University of São Paulo (USP), gfr@hum.ku.dk
**Jennifer D'Souza**, Postdoctoral Researcher at the TIB Leibniz Information Centre for Science and Technology in the Data Sciences and Digital Libraries group, Jennifer.DSouza@tib.eu
**Kim Holmberg**, Senior Researcher at the Research Unit for the Sociology of Education (RUSE) at the University of Turku, kim.holmberg@abo.fi

eer, and Oxley, 2013), importance of a journal to the scientific community (Bergstrom, West, and Wiseman, 2008) or even impact of a journal (Ball, 2017), in which "important journals are those that are highly cited by important journals" (Bohlin et al., 2016). The Eigenfactor algorithm (West, Bergstrom, and Bergstrom, 2010) is based on Eigenvector centrality, which is a commonly used measure to calculate centrality in network analyses (Martin, Zhang, and Newman, 2014).

Bergstrom (2007) describes the approach of ranking journals as similar to the way Google's PageRank algorithm works. Google ranks websites based on the number of hyperlinks between different websites, but all hyperlinks are not considered as equal, as a hyperlink from a website that already receives a significant number of links is more valuable than a hyperlink from a website with only a few links. The Eigenfactor ranks journals in a similar manner by using citations between journals. Bergstrom describes the approach as follows: "We measure the importance of a citation by the influence of the citing journal divided by the total number of citations appearing in that journal" (Bergstrom, 2007). Bergstrom also argues that this approach corrects the differences between journals and disciplines. That is to say, the "Eigenfactor measures the total influence of a journal on the scholarly literature or, comparably, the total value provided by all of the articles published in that journal in a year" (Bergstrom, 2007). Furthermore, Bergstrom developed an article influence rank which "is proportional to the Eigenfactor divided by the number of articles" (Bergstrom, 2007). This rank is comparable to the Journal Impact Factor (Bergstrom, West, and Wiseman, 2008).

Bergstrom (2007) also proposed a way to measure research impact outside the scientific community. This was proposed to be done by calculating references to scholarly articles from a curated list of major newspapers, such as *New York Times*, *The Guardian*, *Wall Street Journal*, *Washington Post*, *London Times*, *Miami Herald*, *Financial Times*, *Le Monde*, *Boston Globe*, and *Los Angeles Times*.

## Role of the Eigenfactor within the Scientific Community

Scientific journals have been an important communication channel for scientific discoveries (Gingras, 2016), ever since the first scientific journal was established in 1665 (Mack, 2015). While there are differences between academic disciplines, such as the social sciences and humanities that have a stronger tradition in publishing books (Hicks, 2005), journals can be found across the range of scientific research. With the introduction of the Internet and the World Wide Web, the importance of scientific journals as a communication and distribution channel has diminished. However, the scientific journal as a publication venue has not changed much since its earliest beginnings (Auer et al., 2018; Wouters et al., 2019). Auer et al. (2018), for example, highlight that journal publications which are mainly based on PDFs could be changed to an interoperable format. This could be done by providing the text in XML (Structured Markup Language). By doing so, the text would provide an improved machine read-ability and linkage between different documents. The final goal with this move could be to interlink this content in a comprehensive knowledge graph. Further initiatives explore the possibility to decentralize the journal publication system by applying blockchain technology (Blocher, Sadeghi, and Sandner, 2019).

Citations have for a long time been considered as recognition of the value of earlier work, i.e. that researchers acknowledge that they have used or found value in the works that they reference. With that, citations have become part of the academic reward system, with highly cited researchers considered to have made a greater impact (Merton, 1973). Citations take, however, a long time to accumulate, as the scientific publishing process can take years. To counter this time delay, journal-based metrics have been developed (Fersht, 2009). The assumption with journal-based impact metrics is that "better" journals have a more rigorous peer review process and that only the "best" research will be published in them. With that, in which journals researchers publish is sometimes even seen as a quality indicator of their work (Chang, McAleer, and Oxley, 2013), which in turn may have consequences on their academic careers (Bohlin et al., 2016; Brembs, Button, and Munafò, 2013) or even generate questionable financial rewards (Quan, Chen, and Shu, 2017). Furthermore, national journal rankings are developed in several countries (Quan, Chen, and Shu, 2017; Huang, 2019). Journal based metrics, such as the Journal Impact Factor, may also be heavily influenced by a small number of articles that receive the majority of citations (Seglen, 1992). Lariviére and Sugimoto (2018), for instance, provided an extensive review of the critique on Journal Impact Factors. Rankings of journals are, thus, a highly-debated topic because they might also affect research assessments (Tüselmann, Sinkovics, and Pishchulov, 2015). On the one hand, journal rankings are oftentimes also accepted by researchers as part of the publishing process (Brembs, Button, and Munafò, 2013), while on the other hand, it has been argued that journals with a higher impact factor seem to be more likely to publish fraudulent work than low-ranked journals (Brembs, Button, and Munafò, 2013; Fang and Casadevall, 2011). Metrics were developed to classify and understand the journal system better (Garfield, 1972), and journal metrics have been developed in several contexts. Furthermore, journal-based metrics can provide a deeper insight into the similarity of journals (D'Souza and Smalheiser, 2014). The first study that tried to develop objective criteria on journals based on citation counts was published in 1927, and focused on the main U.S. chemistry journals for the year 1926. The authors concluded that the majority of journals receive a relatively low number of citations (Gingras, 2016).

As briefly mentioned above, the Eigenfactor was developed as part of a research project at the University of Washington, and the concept is available on a public website. Bergstrom and colleagues tried to serve the needs of various stakeholders, among others the library community, for example, to support librarians' decision-making on journal subscriptions (Kurtz, 2011). One of the goals of the Eigenfactor is to help academic librarians identify the most important journals when deciding which journals to subscribe to. With the constantly increasing subscription prices it is important to know which journals are the most important and that will be used by scholars. This also relates to the fact that with an ever increasing amount of journals (Bohlin et al., 2016; van Gerestein, 2015) a comprehensive overview without rankings and metrics seems impossible. Even if the quality of a journal can only be assessed objectively by human reading of the published articles (Bergstrom, West, and Wiseman, 2008), rankings and metrics to classify journals are a common practice (Bohlin et al., 2016).

Compared to other journal-based metrics, the Eigenfactor has been proposed as an alternative by the San Francisco Declaration on Research Assessment (DORA) (Cagan, 2013). In turn, the Eigenfactor also supports the Initiative for Open Citations (I4OC). The exact extent to which the Eigenfactor is used in the scientific community and research evaluations is unknown. Nevertheless, studies on hiring and tenure promotion provide a glimpse into the use of metrics. Alperin et al. (2018), for example, concluded that metrics, such as the Journal Impact Factor, are used as a measure by hiring and promotion committees in Canada and the United States. The Journal Impact Factor, for example, is used in several decision-making processes in national research systems (Bohlin et al., 2016), and instead of evaluating journals it is also used to evaluate researchers, which is a highly controversial topic (Fersht, 2009; West, Bergstrom, and Bergstrom, 2010; Wouters et al., 2019).

## Critical Perspectives on Journal-based Metrics and Comparison to the Impact Factor

While the Eigenfactor provides some advantages that have been described above, just like any indicator, it does not come without limitations. The Journal Impact Factor was first described in 1972, and is one of the most common journal rankings (Bohlin et al., 2016; Guédon, 2019). It is defined as follows: "The impact factor of a journal in a given year measures the average number of citations to recent articles from articles published in the given year" (Bohlin et al., 2016). The Eigenfactor is also referred to as a rival of the Journal Impact Factor (Reider, 2017) that addresses some of the shortcomings of the former (Tüselmann, Sinkovics, and Pishchulov, 2015). A criticism of the Journal Impact Factor refers to the fact that all citations are assigned the same weight, without taking into account their origin, the journal where the citations occur (Bohlin et al., 2016).

A major difference between the Eigenfactor and the Journal Impact Factor is that the former uses a five-year time window and the latter a two-year window for citations. The broader window should account for citations that appear at a later stage after the research has been published (Bohlin et al., 2016). While a Journal Impact Factor with a five-year time window was also introduced, it seems to be less common than the Journal Impact Factor with a two-year time window (Chang, McAleer, and Oxley, 2013). Another advantage of the Eigenfactor is that self-citations are excluded, which removes score inflations from journal opportunistic self-citations (Bohlin et al., 2016; Chang, McAleer, and Oxley, 2013).

Likewise to the use of any other bibliometric or scientometric indicator, the Eigenfactor should not be used in isolation, and should be supported, for example, by qualitative expert judgements, something that has been emphasised by the Leiden Manifesto for Research Metrics, among others (Hicks et al., 2015). Finally, Bohlin et al. (2016) postulate the most important criterion for evaluating journal-based metrics is the robustness of the method regarding the selection of journals.

## Calculating the Eigenfactor™ Score

The Eigenfactor score is intended to measure the importance of a journal to the scientific community by considering the origin of the incoming citations, and is thought to reflect how frequently an average researcher would access content from that journal. The Eigenfactor for a journal is arrived at by a series of steps (eigenfactor.org). These are elicited below.

First, a five-year cross-citation matrix $Z$ is extracted from the Journal Citation Report (JCR) (clarivate.com).[1]

$Z_{(ij,Y_6)}$ = Citations from journal $j$ in year $Y_6$ to articles published in journal $i$ during the five years $Y_1$ to $Y_5$

For instance, given the 2019 JCR, the entries of the cross-citation matrix would be: $Z_{ij}$ = Citations from journal $j$ in 2019 to articles published in journal $i$ during the 2014 to 2018 five-year period. A longer five-year citation window allows taking into account that certain fields do not have as rapid citation trends as others and only begin a few years after the articles are published. For instance, the average article in a leading cell biology journal might receive 10–30 citations within the two first years after publishing, while, in contrast, the average article in a leading mathematics journal would do very well to receive two citations over the same period. In this regard, measures that only look at citations in the first two years after publication (e.g., Journal Impact Factor) can be misleading (if disciplinary differences are not accounted for).

Note that in $Z$, its diagonal elements are set to 0, thereby omitting journal self-citations. This handles over-inflating journals that engage in the practice of opportunistic self-citation.

In the second step, $Z$ is normalized by the column sums (i.e., by the total number of outgoing citations from each journal) to obtain citation probabilities for each journal column-wise to other journals represented by the matrix rows. The resulting matrix is the column-stochastic matrix $H$, such that

$$H_{ij} = \frac{Z_{(ij,Y_6)}}{\sum_k Z_{(kj,Y_6)}}$$

---

[1] Each year more journals from the Sciences and Social Sciences are indexed in the Journal Citation Report. For the sake of comparison, in 2016, 7611 "source" journals were indexed versus 11,877 in 2019. https://clarivate.com/webofsciencegroup/solutions/journal-citation-reports/ (July 21, 2020).

However, not all the journals listed in $H$ are cited by other journals. These journals will all have 0 entries in their corresponding column $j$. For such journals, an article vector $a$ with entries $a_j$ for each source journal $J$ is computed as follows:

$$a_j = \frac{|J_{Y_1 to\ Y_5}|}{\sum_k |K_{Y_1 to\ Y_5}|}$$

where $|J_{Y_1 to\ Y_5}|$ is the number of articles published by J in the preceding five-year window and the denominator is the number of articles published by all source journals in the JCR over the same five-year window. Thus all journals with no citation links are uniformly populated with $a$, transforming $H$ into $H'$.

Third, a stochastic traversal matrix P is defined following Google's Page-Rank approach, as follows:

$$P = \alpha H' + (1 - \alpha)\alpha.e^T$$

Here, $e^T$ is a row vector of 1s, where $T$ is the transpose function, and thus $A = \alpha.e^T$ is a matrix with identical columns each equal to the article vector $a$.

Under a stochastic process interpretation,[2] the traversal matrix $P$ defines a random walk on the journal citation network that is either a transition with probability $\alpha$ weighted by the entries in $H'$, i.e. the journal citation probabilities, or is a jump to an arbitrary journal with probability $(1 - \alpha)$ weighted by the entries in a, i.e. the proportion of articles published by each journal. Note that without $\alpha$ the traversal will be confined only to the nodes with high $H'$ values. Thus $\alpha$ makes allowance for arbitrary citations not contained in the actual data. At each time instant in the random process $P$ modelling a random walk from journal $J$ to journal $K$, the random variables correspond to matrix values based on the intermediate traversals between journals. Additionally, since $P$ possesses the Markov property[3] whereby traversals to $K$ depend only on knowing the present journal $J$ it came from and no prior history, $P$ is a Markov random process.

---

**2** A stochastic or random process is defined as a collection of random variables indexed at unique time instances.
**3** The Markov property, when applied to stochastic processes, restricts the conditional probability distribution of future states to depend only upon the present state, and not on the entire sequence of states that preceded it, thereby limiting the considered state sequence history.

In the fourth and last step, the Eigenfactor score of each journal is computed. Formally, the Eigenfactor score $EF_i$ of journal $i$ is defined as the percentage of the total weighted citations that journal i receives from source journals. Thus the vector of Eigenfactor scores is written as

$$EF = 100 \frac{H\pi^*}{\sum_i [H\pi^*]_i}$$

where vector $\pi^*$ is extracted from the stochastic traversal matrix $P$ as its leading Eigenvector. Under the stochastic process interpretation the $\pi^*$ vector corresponds to the maximum (also known as steady-state) fraction of time spent at each journal represented in $P$. In the Eigenfactor score, this translates as the measure of the journal influence for weighting citations.

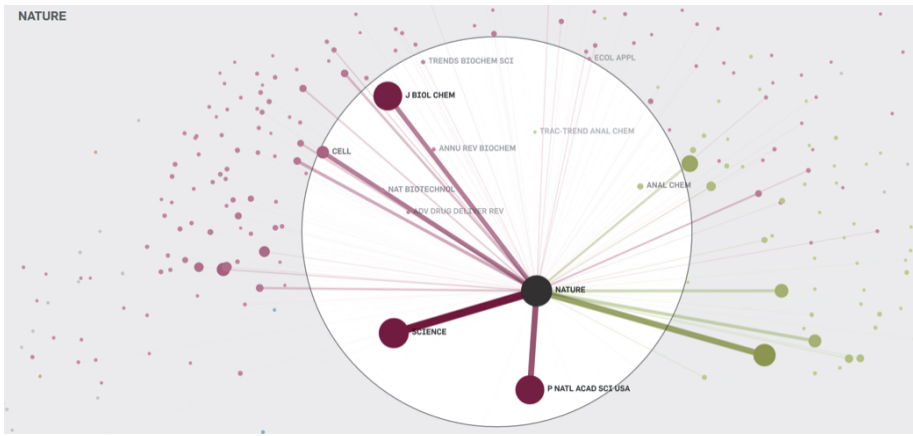

**Figure 1:** A magnified view (within the white circle lens) of a map visualization (well-formed.eigenfactor.org [July 21, 2020]) of Eigenfactor scores (reflected by the node sizes) as a citation network of journals focusing on the journal *Nature* (black node) computed on a subset of journals in the citation data from Thomson Reuters' Journal Citation Reports. For all nodes connected to *Nature*, the edge thicknesses represent the relative amount of citation flow (incoming and outgoing) with respect to it; color-codes correspond to different domains found in the data subset.

Consider in Figure 1 an illustration of the resulting Eigenfactor scores for journals within a citation network. In the figure, the nodes correspond to journals and the nodes sizes reflect their scaled Eigenfactor scores. The calculations presented above recall two aspects involved in computing the Eigenfactor score for the selected journal *Nature* in the figure: (1) citation probabilities from other journals to the journal *Nature* (contained in matrix $H'$ described above); and relying on it, (2) a stochastic traversal pattern defined by $P$ to *Nature*. The figure vaguely depicts this via the edges between *Nature* and other journals and from the edge thicknesses reflecting the citation inflow and outflow. In general, for each journal found in the JCR data, the Eigenfactor score algorithm uses the structure of the entire network to evaluate the importance of each journal, cutting across all disciplines with self-citations excluded. This corresponds to a simple model of research in which readers follow chains of citations as they move from journal to journal. Consequently, journals are considered to be influential if they are cited often by other influential journals.

## Conclusions

As defined by the inventors of the Eigenfactor, "[s]cholarly references join journals together in a vast network of citations" (eigenfactor.org [July 21, 2020]). Given the sheer amount of journals that have emerged over time, this citation data has been measured and analyzed to sort and classify journals.

In this article, one such metric, namely the Eigenfactor, has been presented. Apart from that, the role of this indicator within the scientific community in general has also been addressed. Leveraging the citation data from Clarivate Analytics' Journal Citation Reports (JCR), the Eigenfactor rates journals of science and social science according to the number of incoming citations over a five-year period, with citations from highly ranked journals weighted to make a larger contribution to the Eigenfactor than those from poorly ranked journals via a citation network analysis method inspired from Google's PageRank. Different disciplines have different citing practices and different time scales on which citations occur, therefore the Eigenfactor with its five-year citation window overcomes limitations of its contemporary metric the Journal Impact Factor. This is because the latter kind of metrics with smaller citation time windows can err on the side of assigning higher ratings to journals in disciplines with faster citation patterns rather than creating an allowance for all disciplines and their unique citation patterns.

While metrics might be useful to sort and classify large amounts of data, the concept of a journal's importance also raised criticism. For example, journal-based metrics might have unintended effects on the research system and individual researchers if evaluations are based on metrics without taking into account qualitative expert judgement. There are also initiatives being carried out that try to visualize research outputs beyond journals, and try to acknowledge several forms of impact (e.g., Hauschke, Cartellieri, and Heller, 2018). This article described the Eigenfactor, and mentioned some examples of its role in research systems.

**Acknowledgements**
This chapter was funded by the German Federal Ministry of Education and Research (BMBF) under grant numbers 01PU17019 (Reference Implementation for Open Scientometric Indicators – ROSI) and 16PGF0288 (BMBF Post-Grant-Fund).